\documentclass[aip,apl,reprint, numerical, floatfix]{revtex4-1}
\usepackage{graphicx}
\usepackage{amssymb}
\usepackage{amsmath}
\renewcommand\t[1]{\text{#1}}             % ``text'' i mathenvironment

\newcommand\eq[1]{Eq.~\ref{eq:#1}}
\newcommand\Eqs[1]{Eqs.~\ref{eq:#1}}

\newcommand\cf{\emph{cf.} }

\newcommand\A{\text{A}}

\newcommand\V{\text{V}}

\renewcommand\O{\text{O}}
\renewcommand\H{\text{H}}
\newcommand\VO{{\text{V}_\O^{\cdot\cdot}}}
\newcommand\OH{{\text{OH}_\O^{\cdot}}}

\newcommand\HHO{{\H_2 \text{O}}}

\newcommand\g{\text{g}}
\newcommand\f{\text{f}}
\renewcommand\c{\text{c}}

\renewcommand\i{\text{i}}
\newcommand\hydr{\text{hydr}}

\newcommand\kB{{k_\text{B}}}

\newcommand\comment[1]{}

\newcommand\eqenv[2]{\begin{equation}\label{eq:#1}#2\end{equation}}

\draft % marks overfull lines with a black rule on the right
\begin{document}
\begin{center}
\begin{minipage}[t]{\textwidth}
Copyright 2012 by American Institute of Physics. The following article appeared in Appl. Phys. Lett. \textbf{100}, 061903 (2012); doi: 10.1063/1.3681169 and may be found at http://dx.doi.org/10.1063/1.3681169
\end{minipage}
\end{center}

%\title{Oxygen vacancy segregation and space-charge layers in grain boundaries of dry and hydrated %BaZrO$_\mathbf{3}$} %Title of paper
\title{Oxygen vacancy segregation and space-charge effects in grain boundaries of dry and hydrated BaZrO$_\mathbf{3}$}



\author{B. Joakim Nyman}
\email[]{joakim.nyman@chalmers.se}
%\thanks{}
\author{Edit E. Helgee}
\email[]{edit@chalmers.se}
\author{G\"oran Wahnstr\"om}
\email[]{Corresponding author: goran.wahnstrom@chalmers.se}
\affiliation{Dep. of Applied Physics, Chalmers University of Technology, G\"oteborg, SE-412 96, Sweden}


\date{\today}

\begin{abstract}
A space-charge model is applied to describe the equilibrium effects of segregation of double-donor oxygen
vacancies to grain boundaries in dry and wet acceptor-doped samples of the perovskite oxide BaZrO$_3$.
The grain boundary core vacancy concentrations and electrostatic potential barriers
resulting from different vacancy segregation
energies are evaluated. Density-functional calculations on vacancy segregation to the mirror-symmetric
$\Sigma 3$ (112) [$\bar{1}$10] tilt grain boundary are also presented.
Our results indicate that oxygen vacancy
segregation can be responsible for the low grain boundary proton conductivity in BaZrO$_3$ reported
in the literature.
\end{abstract}


\maketitle %\maketitle must follow title, authors, abstract and \pacs

%%%%%%%%%%%%%%%%%%%%%%%%%%%%%%%%%%%%%%%%%%%%%% Body of paper goes here. Use proper sectioning commands.
While the benefits of ceramic materials make solid oxide proton conductors
like barium zirconate (BaZrO$_3$, BZO) desirable for use as electrolytes
in electrochemical devices such as hydrogen fuel cells, limitations in proton
conductivity have thus far prevented successful implementation.~\cite{POfSOFC2009,fabbri2010} It has
become apparent that boundaries between grains in the material are the
prime source of inhibited overall proton conductivity.~\cite{kreuer2003,bohn2000,pergolesi2010}

Experiments have established that the grain boundary (GB) proton resistivity is an intrinsic effect,
not caused by the segregation of secondary phases at the GB.\cite{iguchi2010,dahl2011} Two
different explanations have instead been put forward~\cite{park2009}, in which the GB effects are believed to
originate in either a structural distortion in the GB region~\cite{kreuer2003}, or in the appearance
of positively charged GBs, caused by a change in chemical composition and leading to Schottky barriers and
the depletion of mobile protonic charge carriers.\cite{iguchi2010,kjolseth2010,chen2011}
At present the latter explanation model is the predominating one~\cite{iguchi2010,kjolseth2010,chen2011},
but the details of the GB cores are neither well understood nor sufficiently explored.

In this letter we present an investigation concerning the equilibrium segregation of
oxygen vacancies to GBs in dry and hydrated acceptor-doped BZO and the
electrostatic potential barrier such segregation causes. Density-functional theory (DFT)
is used to evaluate the energy of vacancy segregation to a $\Sigma 3$
tilt GB and a space-charge model is applied to describe the equilibrium barrier
and core vacancy concentration resulting from different segregation
energies. We show that segregation energies on the order that we find can cause potential
barriers consistent with experimental findings on GB proton conductivity in BZO.

The system is modeled using the periodic supercell technique and
all DFT calculations are performed within the plane-wave
approach as implemented in the Vienna ab-initio simulation package
(VASP)~\cite{Kresse1993,Kresse1996}. Electron-ion interactions are described
by the projector augmented wave method~\cite{Blochl1994} and a generalized gradient
approximation (GGA) exchange-correlation functional due to Perdew and Wang (PW91)~\cite{Wang1991} is used.
More computational details have been reported elsewhere.~\cite{sundell2006i}
In the present work all calculations are performed non-spin-polarized with a plane-wave cutoff of 400~eV
(constant volume) or 520~eV (volume relaxations), and a $4\times 2\times 1$ Monkhorst-Pack $k$-point
grid is used to sample the Brillouin zone.

Theoretical and experimental work on the currently considered $\Sigma 3$ (112) [$\bar{1}$10]
tilt GB has previously been
performed on strontium titanate~\cite{dudeck2010} and this particular GB is well suited for DFT calculations.
It consists of alternating BaZrO and O$_2$
planes and here we consider the mirror-symmetric BaZrO terminated case, which
has the lowest energy. The interplanar distance is \mbox{$d=a_0/\sqrt{24}=0.87$~\AA},
with a computed value of the bulk lattice constant \mbox{$a_0=4.25$~\AA}.
Our supercell has the dimensions ($\sqrt{3}\,a_0,2\sqrt{2}\,a_0,3\sqrt{6}\,a_0$) in the directions
([11$\bar{1}$], [$\bar{1}$10], [112]) and consists of 180 atoms with the distance \mbox{$18d=15.6$~\AA}
between the periodically repeated GBs.

\begin{figure}[tb]
  \centering
  \includegraphics[trim=0mm 0mm 0mm 0mm,clip,width=8.5cm]{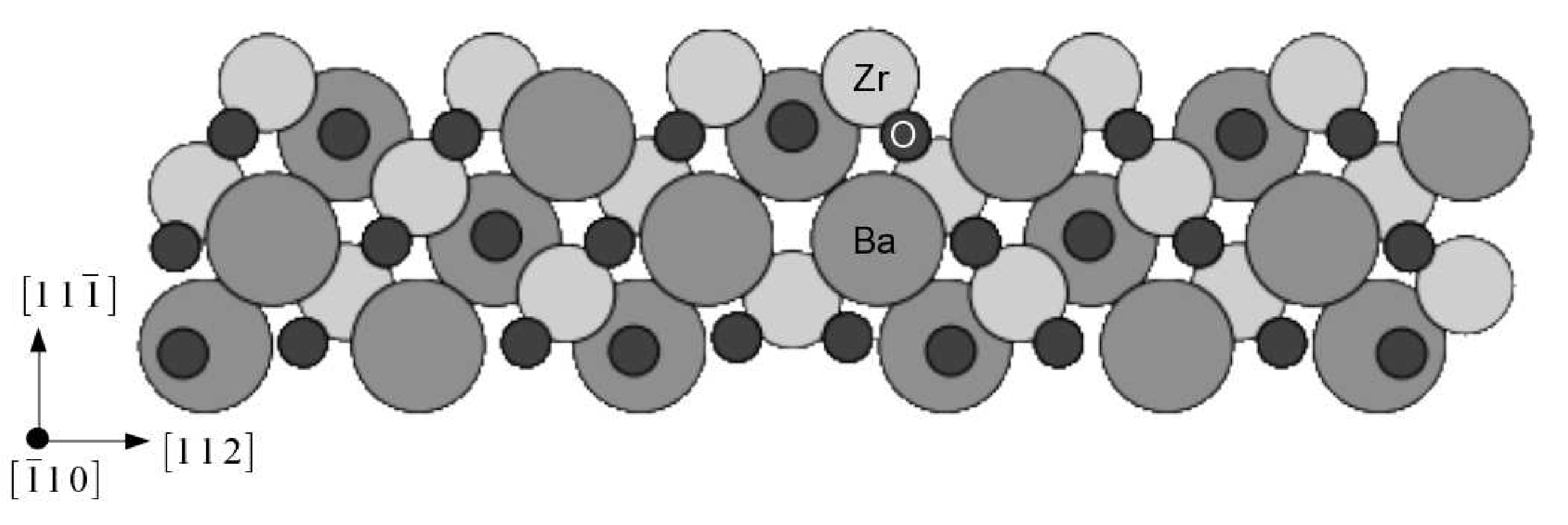}
  	\caption{\label{fig:symgrain}
  	The relaxed structure of the mirror-symmetric $\Sigma$3 (112) [$\bar{1}$10] tilt grain boundary.}
\end{figure}

To begin with, the system is optimized by relaxing the atomic structure, resulting in the configuration
shown in Fig.~1. The GB expands 0.135~\AA\ in the perpendicular direction and the calculated
GB energy\cite{grain-boundary-energy} is \mbox{$\sigma_\text{GB}=0.78$~J/m$^2$}.
Next, an oxygen vacancy is introduced. Since double-donor vacancies are dominating in acceptor
doped BZO, we consider the \mbox{$q=+2$} charge state only.
The vacancy formation energy $\Delta E^\f_\t{GB}$ is determined as function of vacancy position,
both with and without relaxing the structure. As reference, the vacancy formation energy for the bulk
system $\Delta E^\f_\t{bulk}$ with the same size of
the supercell is also determined. The segregation energy is then obtained by taking the difference
\mbox{$E_\t{seg} = \Delta E^\f_\t{GB}-\Delta E^\f_\t{bulk}$}.
Note that the error due to interaction between charged defect images in the periodic supercell
calculations is to a large degree cancelled for the quantity $E_\t{seg}$.
\begin{figure}[tb]
  \centering % trim= l b r t
  \includegraphics[trim=0mm 0mm 0mm 0mm,clip,width=8.5cm]{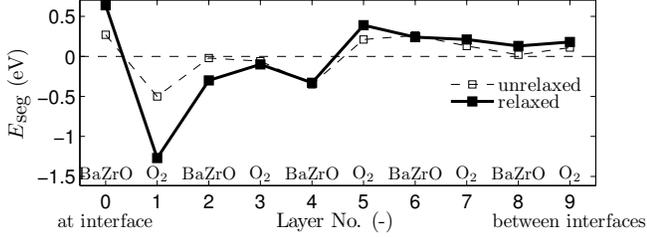}
  	\caption{\label{fig:segregation}
  		Segregation energies of oxygen vacancies in the $\Sigma$3 (112) [$\bar{1}$10] tilt grain boundary as
		function of the location of the vacancy, calculated using DFT.
  	}
\end{figure}

We find a significant segregation tendency to layer 1 of the GB, with an energy of \mbox{$E_\t{seg}=-1.25$~eV}
(\cf Fig.~2). A large part of this energy gain derives from ionic relaxation during vacancy formation.
Referring to Fig.~1, the
high stability of vacancies in layer 1 is due to the close proximity of the pair of oxygen ions on
either side of the interface and the electrostatic repulsion this causes. As one of the oxygens in the
pair is removed, the other ion relaxes considerably, to a position in the symmetry plane.

Having established the existence of low-energy vacancy positions close to the GB interface,
we  would now like to evaluate the corresponding
equilibrium vacancy segregation and electrostatic potential barrier as function of temperature.
To do so, we apply a space charge model~\cite{maier1995ionic,desouza2009,waser2006,kim2003space} with an ideal depletion approximation
to two different cases: i) dry BZO, where a concentration of single-acceptor dopants are perfectly compensated
for by double-donor vacancies, and ii) wet BZO, where dissociative absorption of water molecules from a humid
atmosphere is taken into account.

In both cases we consider a one-dimensional model, with three regions of constant defect concentration:
i) the GB core \mbox{($0<x<x_0$)}, where an increased concentration $c_\V^\c$ of oxygen vacancies is
expected and the proton concentration is assumed to vanish, ii) a compensating space charge layer
(SCL) \mbox{($x_0<x<x^*$)}, depleted  of oxygen vacancies and protons, and finally iii) the neutral
grain interior \mbox{($x>x^*$)} with vacancy concentration $c_\V^\i$ and in the wet case proton
concentration $c_\t{OH}^\i$. The dopant concentration $c_A$ is taken to be fixed and equal throughout all
regions and a
concentration corresponding to occupation of 10\% of the Zr sites is used in all calculations.
Motivated by the high stability of oxygen vacancies in layer 1 (\cf Fig.~2), we define a core region
enclosing that layer and denote the segregation energy for those vacancies with $E_\t{seg}^\c$.
Vacancies in layers 2,3,4,$\ldots$ are treated as having zero segregation energy and the symmetry plane, layer 0,
is assumed void of vacancies. For the core half-width we use the value \mbox{$x_0 = 2$~\AA}.
With this description the density of active oxygen vacancy sites in the core becomes about half that of the grain
interior: $N^{\c}=0.579\,N^{\i}$, where $N^{\i}=3/a_0^3=0.039$~\AA$^{-3}$.

We begin with the dry case and consider electrochemical equilibrium between oxygen vacancies in the
GB core and grain interior for an ideal solution~\cite{maier2004}:
\eqenv{equilibrium}{%
	E_\t{seg}^\c + 2\,e\,\Delta\phi^\c + \kB T\,\ln\frac{c_\V^\c}{N^\c-c_\V^c} =
	\kB T\,\ln\frac{c_\V^\i}{N^\i-c_\V^\i}.
}
The electrostatic potential difference between core and interior, $\Delta\phi^\c\equiv \phi(x_0) - \phi(x^*)$,
is obtained by solving Poisson's
equation in the SCL with the boundary conditions \mbox{$\phi(x^*)=\phi'(x^*)=0$}. For the dielectric constant
we use the value \mbox{$\epsilon=32\,\epsilon_0$\cite{LB36a1}} and the size of the SCL is determined from
charge compensation with respect to the core: \mbox{$(2 e c_\V^\c - e c_A)\,x_0 = e c_A\,(x^* - x_0)$}, or
\mbox{$x^* = (2 c_\V^\c / c_A)\,x_0$}. It follows that the present model is applicable for $c_\V^\c>c_A/2$ only.
The solution for the potential difference is:
\eqenv{potential}{%
	\Delta\phi^\c \equiv \phi(x_0) - \phi(x^*)
    = \frac{e}{2\epsilon} \frac{(2c_\V^\c - c_A)^2}{c_A}\,x_0^2.
}
\Eqs{equilibrium} and \ref{eq:potential} can now be solved iteratively to obtain $c_\V^\c$ and $\Delta\phi^\c$ as
function of temperature. Using the neutrality condition, the grain interior oxygen vacancy concentration is in the present, dry, case given by: \mbox{$c_\V^\i = c_A/2$}. Our numerical results are shown in Fig.~3 as dashed
lines for a few different values of the segregation energy $E_\t{seg}^\c$.

Next, the wet case. Hydration of the grain interior is modeled by the hydration
reaction (Kr\"oger-Vink notation):
\eqenv{hydration}{
	\HHO(\g) + \VO + \O^\times_\O \longleftrightarrow 2\,\OH.
}
Experimental values from the literature~\cite{kreuer2003} are assigned to the hydration enthalpy
\mbox{$\Delta H^0_\hydr=-0.82$~eV} and entropy \mbox{$\Delta S^0_\hydr=-0.92$~meV/K}
and a water partial pressure of \mbox{$p_\HHO=0.025$~bar} is used in the calculations. The interior hydroxide
concentration $c_\t{OH}^\i$ is obtained from the law of mass action, charge neutrality and site
restriction~\cite{kreuer2003}:
\eqenv{hydroxide_conc}{%
	\frac{c_\t{OH}^\i}{N^\i} = \frac{\kappa}{\kappa-4}\left[ 1-\sqrt{1-\frac{\kappa-4}{\kappa}\left(2\frac{c_A}{N^\i}-
\left(\frac{c_A}{{N^\i}}\right)^2\right)} \right],
}
where \mbox{$\kappa = p_\HHO\,K$} and the equilibrium constant
\mbox{$K=\exp(\Delta S^0_\hydr/\kB)\exp(-\Delta H^0_\hydr/\kB T)$}. In the middle panel of
Fig.~3 we show the result for the interior hydroxide concentration
$c_\t{OH}^\i$, indicating the transition between wet and dry grain interior around 900 K.

We are now in the position to determine the concentration of core vacancies $c_\V^\c$ and the
electrostatic barrier
$\Delta\phi^\c$ as function of temperature in hydrated samples. In this case
\mbox{$c_\V^\i = (c_A-c_\t{OH}^\i)/2$} and
\mbox{$N^\i-c_\V^\i$} in \eq{equilibrium} is replaced by \mbox{$N^\i-c_\t{OH}^\i-c_\V^\i$}.
\Eqs{equilibrium} and \ref{eq:potential} are solved together with the expression
for the hydroxide concentration in \eq{hydroxide_conc} and our final results for $c_\V^\c$ and $\Delta\phi^\c$ in
hydrated samples are shown as solid lines in Fig.~3.

%%%%%%%%%%%%%%%%%%%%%%%%%%%%%%%%%%%%%%%%%%%%%%%%%%%%%%%%%%%%%%%%%%%%%%%%%%%%%%%%%%%%%%%%%%%%%%%%%%%%%%%

%The results of our space charge model for different segregation energies can be seen in \fig{spacecharge}.
\begin{figure}[tb]
  \centering % trim= l b r t
  \includegraphics[trim=0mm 0mm 0mm 0mm,clip,width=8.5cm]{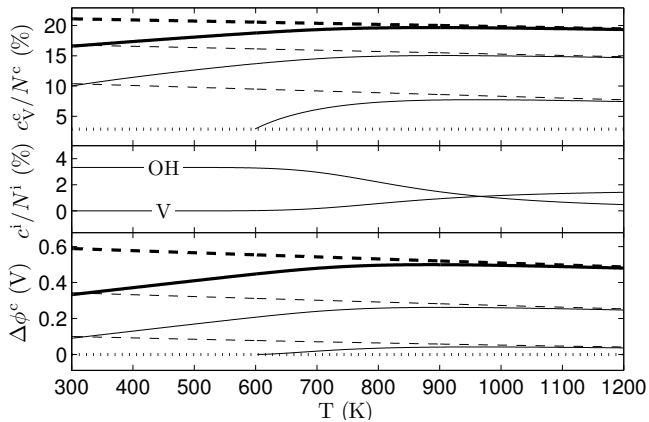}
  	\caption{\label{fig:spacecharge}
  		Core vacancy site-concentration and potential barrier
  		for vacancy segregation energies $E_\t{seg}^\c=\{-0.25,-0.75,-1.25\}$~eV, where the thick lines
		correspond to $-1.25$~eV. Dashed/solid lines have been used for the dry/hydrated case and dotted
		lines indicate the bound of the present space-charge model, i.e. $c_\V^\c>c_\A/2$. The middle panel shows interior
		vacancy- and hydroxide concentrations for the hydrated case.}
\end{figure}%
It is clear from Fig.~3 that, according to our model, segregation energies on the order that
we find in the $\Sigma 3$ tilt GB can lead to significant electrostatic potential barriers
as a result of oxygen vacancy aggregation at the GB. In dry BZO at 600~K, 21\% of the
oxygen sites having a vacancy segregation energy of $-1.25$~eV in the GB core are predicted to be
vacant, giving rise to an electrostatic potential barrier of 0.55~V.
In the hydrated case, vacancies persist at the GB to the degree
that, although the interior vacancies have been annihilated at 600~K, 19\% of the core sites are
still vacant and the barrier is 0.45~V. Even if the sample is cooled to room temperature, the number
(assuming thermodynamic equilibrium) only drops to 17\% and a potential barrier of 0.33~V is still present.

In recent work, Kj{\o}lseth et al~\cite{kjolseth2010} used a space-charge model to evaluate the
potential barrier corresponding to experimental results on grain interior and grain boundary proton
conductivity in Y-doped BZO~\cite{kjolseth2010,duval2007,babilo2007}. A barrier on the order
0.4--0.6~V was found to match measurements made at 200--300$^\circ$C. This agrees relatively well with
our results, where a barrier of 0.4--0.45~V is seen for \mbox{$E_\t{seg}^\c=-1.25$~eV} in this temperature
range. Similar potential barriers fitted to experimental conductivity data were also found
by Chen~et~al~\cite{chen2011} for 10~mol\% Y-doped samples~\cite{chen2011,kjolseth2010,duval2007,iguchi2010}
and by Iguchi~et~al~\cite{iguchi2010}.
Interesting to point out is the significant difference in potential barrier
calculated from conductivity measurements of Duval~\cite{duval2007}, of BZO annealed at high temperature and
of conventionally sintered samples. Chen~et~al calculated a barrier of 0.77~V in the conventionally sintered
material, while the calculations of Kj{\o}lseth~et~al show a barrier of 0.46~V in the annealed sample.

%This
%difference, along with the generally large spread in grain boundary conductivity and estimated potentials
%seen depending on materials processing techniques, might be indicative of differing grain boundary
%structures. Difficulties to obtain equilibrated interfaces between grains in BZO is consistent with the
%refractory nature of BZO and also explains why our results for a fully optimized grain boundary structure
%fall in the lower end of the estimated potential range.

When comparing our results to barriers estimated from experimental data it is important to keep in
mind that our results are based on DFT calculation on one single grain boundary structure, while
experimental results  correspond to an average effect in polycrystalline samples.
Theoretical investigations of additional GBs are therefore
warranted to conclude whether the magnitude of the present oxygen
vacancy segregation energy is a general trait of GBs in BZO. Furthermore an aggregation of acceptor
dopants at GBs has been observed in BZO~\cite{gross2001bazr0.85me0.15o2.925,kjolseth2010,iguchi2010},
an effect which we do not take into account and which would tend to diminish the positive charge due to
oxygen vacancies. In addition, there
is reason to investigate the segregation energy of protons in GBs, particularly considering that
the stability of protons in solid oxides has been seen to increase on oxygen sites where vacancies
are also increasingly stable~\cite{bjorheim2010,nyman2011}.

In conclusion, we present DFT calculations of oxygen vacancy segregation to a $\Sigma 3$ tilt
grain boundary in BZO. Using a space charge model we demonstrate that at 600 K, where the sample is fully
hydrated in a wet atmosphere, the oxygen vacancies in the grain boundary core are still present with a site
concentration of about 20\%. This corresponds to a Schottky barrier height of 0.45~V and compares well
with recent experimental data~\cite{kjolseth2010,chen2011,duval2007,babilo2007,iguchi2010}.
Our results indicate that oxygen vacancy
segregation can be responsible for the low grain boundary proton conductivity in BZO reported
in the literature.

%\emph{Since this paper was submitted for publication further relevant work on space charge and blocking effects %at grain boundaries of BZO has been published~\cite{shirpour????evidence,iguchi2011,deSouza2011}.}
Since this paper was submitted for publication further relevant work on space charge and blocking effects at grain boundaries of BZO has been published~\cite{shirpour????evidence,iguchi2011,deSouza2011}.

We would like to acknowledge Profs Joachim Maier and Truls Norby for useful conversations.
The computations have been performed on SNIC resources and the Swedish Energy Agency is acknowledged for financial
support.

%%%%%%%%%%%%%%%%%%%%%%%%%%%%%%%%%%%%%%%%%%%%%%%%%%%%%%%%%% Create the reference section using BibTeX:
%\bibliographystyle{numerical}
%\bibliography{REFERENSER_JN.bib}

\begin{thebibliography}{24}%
\makeatletter
\providecommand \@ifxundefined [1]{%
 \@ifx{#1\undefined}
}%
\providecommand \@ifnum [1]{%
 \ifnum #1\expandafter \@firstoftwo
 \else \expandafter \@secondoftwo
 \fi
}%
\providecommand \@ifx [1]{%
 \ifx #1\expandafter \@firstoftwo
 \else \expandafter \@secondoftwo
 \fi
}%
\providecommand \natexlab [1]{#1}%
\providecommand \enquote  [1]{``#1''}%
\providecommand \bibnamefont  [1]{#1}%
\providecommand \bibfnamefont [1]{#1}%
\providecommand \citenamefont [1]{#1}%
\providecommand \href@noop [0]{\@secondoftwo}%
\providecommand \href [0]{\begingroup \@sanitize@url \@href}%
\providecommand \@href[1]{\@@startlink{#1}\@@href}%
\providecommand \@@href[1]{\endgroup#1\@@endlink}%
\providecommand \@sanitize@url [0]{\catcode `\\12\catcode `\$12\catcode
  `\&12\catcode `\#12\catcode `\^12\catcode `\_12\catcode `\%12\relax}%
\providecommand \@@startlink[1]{}%
\providecommand \@@endlink[0]{}%
\providecommand \url  [0]{\begingroup\@sanitize@url \@url }%
\providecommand \@url [1]{\endgroup\@href {#1}{\urlprefix }}%
\providecommand \urlprefix  [0]{URL }%
\providecommand \Eprint [0]{\href }%
\providecommand \doibase [0]{http://dx.doi.org/}%
\providecommand \selectlanguage [0]{\@gobble}%
\providecommand \bibinfo  [0]{\@secondoftwo}%
\providecommand \bibfield  [0]{\@secondoftwo}%
\providecommand \translation [1]{[#1]}%
\providecommand \BibitemOpen [0]{}%
\providecommand \bibitemStop [0]{}%
\providecommand \bibitemNoStop [0]{.\EOS\space}%
\providecommand \EOS [0]{\spacefactor3000\relax}%
\providecommand \BibitemShut  [1]{\csname bibitem#1\endcsname}%
\let\auto@bib@innerbib\@empty
%</preamble>
\bibitem [{\citenamefont {Ishihara}(2009)}]{POfSOFC2009}%
  \BibitemOpen
  \bibinfo {editor} {\bibfnamefont {T.}~\bibnamefont {Ishihara}},\ ed.,\
  \href@noop {} {\emph {\bibinfo {title} {Perovskite Oxide for Solid Oxide Fuel
  Cells}}}\ (\bibinfo  {publisher} {Springer},\ \bibinfo {address} {Dordrecht},\
  \bibinfo {year} {2009})\BibitemShut {NoStop}%
\bibitem [{\citenamefont {Fabbri}, \citenamefont {Pergolesi},\ and\
  \citenamefont {Traversa}(2010)}]{fabbri2010}%
  \BibitemOpen
  \bibfield  {author} {\bibinfo {author} {\bibfnamefont {E.}~\bibnamefont
  {Fabbri}}, \bibinfo {author} {\bibfnamefont {D.}~\bibnamefont {Pergolesi}}, \
  and\ \bibinfo {author} {\bibfnamefont {E.}~\bibnamefont {Traversa}},\
  }\href@noop {} {\bibfield  {journal} {\bibinfo  {journal} {Chem.~Soc.~Rev.}\
  }\textbf {\bibinfo {volume} {39}},\ \bibinfo {pages} {4355} (\bibinfo {year}
  {2010})}\BibitemShut {NoStop}%
\bibitem [{\citenamefont {Kreuer}(2003)}]{kreuer2003}%
  \BibitemOpen
  \bibfield  {author} {\bibinfo {author} {\bibfnamefont {K.~D.}\ \bibnamefont
  {Kreuer}},\ }\href@noop {} {\bibfield  {journal} {\bibinfo  {journal} {Annu.
  Rev. Mater. Res.}\ }\textbf {\bibinfo {volume} {33}},\ \bibinfo {pages} {333}
  (\bibinfo {year} {2003})}\BibitemShut {NoStop}%
\bibitem [{\citenamefont {Bohn}\ and\ \citenamefont
  {Schober}(2000)}]{bohn2000}%
  \BibitemOpen
  \bibfield  {author} {\bibinfo {author} {\bibfnamefont {H.~G.}\ \bibnamefont
  {Bohn}}\ and\ \bibinfo {author} {\bibfnamefont {T.}~\bibnamefont {Schober}},\
  }\href@noop {} {\bibfield  {journal} {\bibinfo  {journal} {J. Am. Ceram.
  Soc.}\ }\textbf {\bibinfo {volume} {83}},\ \bibinfo {pages} {768} (\bibinfo
  {year} {2000})}\BibitemShut {NoStop}%
\bibitem [{\citenamefont {Pergolesi}\ \emph {et~al.}(2010)\citenamefont
  {Pergolesi}, \citenamefont {Fabbri}, \citenamefont {D´Epifanio},
  \citenamefont {Bartolomeo}, \citenamefont {Tebano}, \citenamefont {Sanna},
  \citenamefont {Licocca}, \citenamefont {Balestrino},\ and\ \citenamefont
  {Traversa}}]{pergolesi2010}%
  \BibitemOpen
  \bibfield  {author} {\bibinfo {author} {\bibfnamefont {D.}~\bibnamefont
  {Pergolesi}}, \bibinfo {author} {\bibfnamefont {E.}~\bibnamefont {Fabbri}},
  \bibinfo {author} {\bibfnamefont {A.}~\bibnamefont {D´Epifanio}}, \bibinfo
  {author} {\bibfnamefont {E.~D.}\ \bibnamefont {Bartolomeo}}, \bibinfo
  {author} {\bibfnamefont {A.}~\bibnamefont {Tebano}}, \bibinfo {author}
  {\bibfnamefont {S.}~\bibnamefont {Sanna}}, \bibinfo {author} {\bibfnamefont
  {S.}~\bibnamefont {Licocca}}, \bibinfo {author} {\bibfnamefont
  {G.}~\bibnamefont {Balestrino}}, \ and\ \bibinfo {author} {\bibfnamefont
  {E.}~\bibnamefont {Traversa}},\ }\href@noop {} {\bibfield  {journal}
  {\bibinfo  {journal} {Nat.~Mater.}\ }\textbf {\bibinfo {volume} {9}},\
  \bibinfo {pages} {846} (\bibinfo {year} {2010})}\BibitemShut {NoStop}%
\bibitem [{\citenamefont {Iguchi}, \citenamefont {Sata},\ and\ \citenamefont
  {Yugami}(2010)}]{iguchi2010}%
  \BibitemOpen
  \bibfield  {author} {\bibinfo {author} {\bibfnamefont {F.}~\bibnamefont
  {Iguchi}}, \bibinfo {author} {\bibfnamefont {N.}~\bibnamefont {Sata}}, \ and\
  \bibinfo {author} {\bibfnamefont {H.}~\bibnamefont {Yugami}},\ }\href@noop {}
  {\bibfield  {journal} {\bibinfo  {journal} {J.~Mater.~Chem.}\ }\textbf
  {\bibinfo {volume} {20}},\ \bibinfo {pages} {6265} (\bibinfo {year}
  {2010})}\BibitemShut {NoStop}%
\bibitem [{\citenamefont {Dahl}\ \emph {et~al.}(2011)\citenamefont {Dahl},
  \citenamefont {Lein}, \citenamefont {Yu}, \citenamefont {Tolchard},
  \citenamefont {Grande}, \citenamefont {Einarsrud}, \citenamefont
  {Kj{\o}lseth}, \citenamefont {Norby},\ and\ \citenamefont
  {Haugsrud}}]{dahl2011}%
  \BibitemOpen
  \bibfield  {author} {\bibinfo {author} {\bibfnamefont {P.}~\bibnamefont
  {Dahl}}, \bibinfo {author} {\bibfnamefont {H.}~\bibnamefont {Lein}}, \bibinfo
  {author} {\bibfnamefont {Y.}~\bibnamefont {Yu}}, \bibinfo {author}
  {\bibfnamefont {J.}~\bibnamefont {Tolchard}}, \bibinfo {author}
  {\bibfnamefont {T.}~\bibnamefont {Grande}}, \bibinfo {author} {\bibfnamefont
  {M.}~\bibnamefont {Einarsrud}}, \bibinfo {author} {\bibfnamefont
  {C.}~\bibnamefont {Kj{\o}lseth}}, \bibinfo {author} {\bibfnamefont
  {T.}~\bibnamefont {Norby}}, \ and\ \bibinfo {author} {\bibfnamefont
  {R.}~\bibnamefont {Haugsrud}},\ }\href@noop {} {\bibfield  {journal}
  {\bibinfo  {journal} {Solid State Ionics}\ }\textbf {\bibinfo {volume}
  {182}},\ \bibinfo {pages} {32} (\bibinfo {year} {2011})}\BibitemShut
  {NoStop}%
\bibitem [{\citenamefont {Park}\ \emph {et~al.}(2009)\citenamefont {Park},
  \citenamefont {Kwak}, \citenamefont {Lee}, \citenamefont {Lee},\ and\
  \citenamefont {Lee}}]{park2009}%
  \BibitemOpen
  \bibfield  {author} {\bibinfo {author} {\bibfnamefont {H.}~\bibnamefont
  {Park}}, \bibinfo {author} {\bibfnamefont {C.}~\bibnamefont {Kwak}}, \bibinfo
  {author} {\bibfnamefont {K.}~\bibnamefont {Lee}}, \bibinfo {author}
  {\bibfnamefont {S.}~\bibnamefont {Lee}}, \ and\ \bibinfo {author}
  {\bibfnamefont {E.}~\bibnamefont {Lee}},\ }\href@noop {} {\bibfield
  {journal} {\bibinfo  {journal} {J.~Eur.~Ceram.~Soc.}\ }\textbf {\bibinfo
  {volume} {29}},\ \bibinfo {pages} {2429} (\bibinfo {year}
  {2009})}\BibitemShut {NoStop}%
\bibitem [{\citenamefont {Kj{\o}lseth}\ \emph {et~al.}(2010)\citenamefont
  {Kj{\o}lseth}, \citenamefont {Fjeld}, \citenamefont {Prytz}, \citenamefont
  {Dahl}, \citenamefont {Estournes}, \citenamefont {Haugsrud},\ and\
  \citenamefont {Norby}}]{kjolseth2010}%
  \BibitemOpen
  \bibfield  {author} {\bibinfo {author} {\bibfnamefont {C.}~\bibnamefont
  {Kj{\o}lseth}}, \bibinfo {author} {\bibfnamefont {H.}~\bibnamefont {Fjeld}},
  \bibinfo {author} {\bibfnamefont {{\O}.}~\bibnamefont {Prytz}}, \bibinfo
  {author} {\bibfnamefont {P.~I.}\ \bibnamefont {Dahl}}, \bibinfo {author}
  {\bibfnamefont {C.}~\bibnamefont {Estournes}}, \bibinfo {author}
  {\bibfnamefont {R.}~\bibnamefont {Haugsrud}}, \ and\ \bibinfo {author}
  {\bibfnamefont {T.}~\bibnamefont {Norby}},\ }\href@noop {} {\bibfield
  {journal} {\bibinfo  {journal} {Solid State Ionics}\ }\textbf {\bibinfo
  {volume} {181}},\ \bibinfo {pages} {268} (\bibinfo {year}
  {2010})}\BibitemShut {NoStop}%
\bibitem [{\citenamefont {Chen}, \citenamefont {Danel},\ and\ \citenamefont
  {Kim}(2011)}]{chen2011}%
  \BibitemOpen
  \bibfield  {author} {\bibinfo {author} {\bibfnamefont {C.}~\bibnamefont
  {Chen}}, \bibinfo {author} {\bibfnamefont {C.}~\bibnamefont {Danel}}, \ and\
  \bibinfo {author} {\bibfnamefont {S.}~\bibnamefont {Kim}},\ }\href@noop {}
  {\bibfield  {journal} {\bibinfo  {journal} {J.~Mater.~Chem.}\ }\textbf
  {\bibinfo {volume} {21}},\ \bibinfo {pages} {5435} (\bibinfo {year}
  {2011})}\BibitemShut {NoStop}%
\bibitem [{\citenamefont {Kresse}\ and\ \citenamefont
  {Hafner}(1993)}]{Kresse1993}%
  \BibitemOpen
  \bibfield  {author} {\bibinfo {author} {\bibfnamefont {G.}~\bibnamefont
  {Kresse}}\ and\ \bibinfo {author} {\bibfnamefont {J.}~\bibnamefont
  {Hafner}},\ }\href@noop {} {\bibfield  {journal} {\bibinfo  {journal} {Phys.
  Rev. B}\ }\textbf {\bibinfo {volume} {48}},\ \bibinfo {pages} {13115}
  (\bibinfo {year} {1993})}\BibitemShut {NoStop}%
\bibitem [{\citenamefont {Kresse}\ and\ \citenamefont
  {Furthm{\"u}ller}(1996)}]{Kresse1996}%
  \BibitemOpen
  \bibfield  {author} {\bibinfo {author} {\bibfnamefont {G.}~\bibnamefont
  {Kresse}}\ and\ \bibinfo {author} {\bibfnamefont {J.}~\bibnamefont
  {Furthm{\"u}ller}},\ }\href@noop {} {\bibfield  {journal} {\bibinfo
  {journal} {Phys. Rev. B}\ }\textbf {\bibinfo {volume} {54}},\ \bibinfo
  {pages} {11169} (\bibinfo {year} {1996})}\BibitemShut {NoStop}%
\bibitem [{\citenamefont {Bl{\"o}chl}(1994)}]{Blochl1994}%
  \BibitemOpen
  \bibfield  {author} {\bibinfo {author} {\bibfnamefont {P.~E.}\ \bibnamefont
  {Bl{\"o}chl}},\ }\href@noop {} {\bibfield  {journal} {\bibinfo  {journal}
  {Phys. Rev. B}\ }\textbf {\bibinfo {volume} {50}},\ \bibinfo {pages} {17953}
  (\bibinfo {year} {1994})}\BibitemShut {NoStop}%
\bibitem [{\citenamefont {Wang}\ and\ \citenamefont {Perdew}(1991)}]{Wang1991}%
  \BibitemOpen
  \bibfield  {author} {\bibinfo {author} {\bibfnamefont {Y.}~\bibnamefont
  {Wang}}\ and\ \bibinfo {author} {\bibfnamefont {J.~P.}\ \bibnamefont
  {Perdew}},\ }\href@noop {} {\bibfield  {journal} {\bibinfo  {journal} {Phys.
  Rev. B}\ }\textbf {\bibinfo {volume} {44}},\ \bibinfo {pages} {13298}
  (\bibinfo {year} {1991})}\BibitemShut {NoStop}%
\bibitem [{\citenamefont {Sundell}, \citenamefont {Bj{\"o}rketun},\ and\
  \citenamefont {Wahnstr{\"o}m}(2006)}]{sundell2006i}%
  \BibitemOpen
  \bibfield  {author} {\bibinfo {author} {\bibfnamefont {P.~G.}\ \bibnamefont
  {Sundell}}, \bibinfo {author} {\bibfnamefont {M.~E.}\ \bibnamefont
  {Bj{\"o}rketun}}, \ and\ \bibinfo {author} {\bibfnamefont {G.}~\bibnamefont
  {Wahnstr{\"o}m}},\ }\href@noop {} {\bibfield  {journal} {\bibinfo  {journal}
  {Phys. Rev. B}\ }\textbf {\bibinfo {volume} {73}},\ \bibinfo {pages} {104112}
  (\bibinfo {year} {2006})}\BibitemShut {NoStop}%
\bibitem [{\citenamefont {{K.~J.~Dudeck and {N.~A.}~Benedek and
  {D.~J.~H.}~Cockayne}}(2010)}]{dudeck2010}%
  \BibitemOpen
  \bibfield  {author} {\bibinfo {author} {\bibnamefont {{K.~J.~Dudeck and
  {N.~A.}~Benedek and {D.~J.~H.}~Cockayne}}},\ }\href@noop {} {\bibfield
  {journal} {\bibinfo  {journal} {Phys.~Rev.~B}\ }\textbf {\bibinfo {volume}
  {81}},\ \bibinfo {pages} {134109} (\bibinfo {year} {2010})}\BibitemShut
  {NoStop}%
\bibitem [{gra()}]{grain-boundary-energy}%
  \BibitemOpen
  \href@noop {} {}\bibinfo {note} {The GB energy is defined as
  $\sigma_\text{GB} = (E_\t{GB}-E_\t{bulk})/2A$, where $E_\t{GB}$ and
  $E_\t{bulk}$ are the DFT total energies of the GB and bulk supercell
  structures respectively, and $A$ is the area of the GB interface in the
  supercell.}\BibitemShut {Stop}%
\bibitem [{\citenamefont {{De~Souza}}(2009)}]{desouza2009}%
  \BibitemOpen
  \bibfield  {author} {\bibinfo {author} {\bibfnamefont {R.~A.}\ \bibnamefont
  {{De~Souza}}},\ }\href@noop {} {\bibfield  {journal} {\bibinfo  {journal}
  {Phys. Chem. Chem. Phys.}\ }\textbf {\bibinfo {volume} {11}},\ \bibinfo
  {pages} {9939} (\bibinfo {year} {2009})}\BibitemShut {NoStop}%
\bibitem [{\citenamefont {Guo}\ and\ \citenamefont {Waser}(2006)}]{waser2006}%
  \BibitemOpen
  \bibfield  {author} {\bibinfo {author} {\bibfnamefont {X.}~\bibnamefont
  {Guo}}\ and\ \bibinfo {author} {\bibfnamefont {R.}~\bibnamefont {Waser}},\
  }\href@noop {} {\bibfield  {journal} {\bibinfo  {journal}
  {Prog.~Mater.~Sci.}\ }\textbf {\bibinfo {volume} {51}},\ \bibinfo {pages}
  {151} (\bibinfo {year} {2006})}\BibitemShut {NoStop}%
\bibitem [{\citenamefont {Maier}(1995)}]{maier1995ionic}%
  \BibitemOpen
  \bibfield  {author} {\bibinfo {author} {\bibfnamefont {J.}~\bibnamefont
  {Maier}},\ }\href {\doibase 10.1016/0079-6786(95)00004-E} {\bibfield
  {journal} {\bibinfo  {journal} {Progress in Solid State Chemistry}\ }\textbf
  {\bibinfo {volume} {23}},\ \bibinfo {pages} {171} (\bibinfo {year}
  {1995})}\BibitemShut {NoStop}%
\bibitem [{\citenamefont {Kim}, \citenamefont {Fleig},\ and\ \citenamefont
  {Maier}(2003)}]{kim2003space}%
  \BibitemOpen
  \bibfield  {author} {\bibinfo {author} {\bibfnamefont {S.}~\bibnamefont
  {Kim}}, \bibinfo {author} {\bibfnamefont {J.}~\bibnamefont {Fleig}}, \ and\
  \bibinfo {author} {\bibfnamefont {J.}~\bibnamefont {Maier}},\ }\href
  {\doibase 10.1039/B300170A} {\bibfield  {journal} {\bibinfo  {journal} {Phys.
  Chem. Chem. Phys.}\ }\textbf {\bibinfo {volume} {5}},\ \bibinfo {pages}
  {2268} (\bibinfo {year} {2003})}\BibitemShut {NoStop}%
\bibitem [{\citenamefont {Maier}(2004)}]{maier2004}%
  \BibitemOpen
  \bibfield  {author} {\bibinfo {author} {\bibfnamefont {J.}~\bibnamefont
  {Maier}},\ }\href@noop {} {\emph {\bibinfo {title} {Physical Chemistry of
  Ionic Materials}}}\ (\bibinfo  {publisher} {John Wiley \& Sons},\ \bibinfo
  {address} {Chichester, West Sussex},\ \bibinfo {year} {2004})\BibitemShut
  {NoStop}%
\bibitem [{\citenamefont {Adachi}\ \emph {et~al.}(2001)\citenamefont {Adachi},
  \citenamefont {Akishige}, \citenamefont {Asahi}, \citenamefont {Deguchi},
  \citenamefont {Gesi}, \citenamefont {Hasebe}, \citenamefont {Hikita},
  \citenamefont {Ikeda},\ and\ \citenamefont {Iwata}}]{LB36a1}%
  \BibitemOpen
  \bibfield  {author} {\bibinfo {author} {\bibfnamefont {M.}~\bibnamefont
  {Adachi}}, \bibinfo {author} {\bibfnamefont {Y.}~\bibnamefont {Akishige}},
  \bibinfo {author} {\bibfnamefont {T.}~\bibnamefont {Asahi}}, \bibinfo
  {author} {\bibfnamefont {K.}~\bibnamefont {Deguchi}}, \bibinfo {author}
  {\bibfnamefont {K.}~\bibnamefont {Gesi}}, \bibinfo {author} {\bibfnamefont
  {K.}~\bibnamefont {Hasebe}}, \bibinfo {author} {\bibfnamefont
  {T.}~\bibnamefont {Hikita}}, \bibinfo {author} {\bibfnamefont
  {T.}~\bibnamefont {Ikeda}}, \ and\ \bibinfo {author} {\bibfnamefont
  {Y.}~\bibnamefont {Iwata}},\ }\href@noop {} {\emph {\bibinfo {title}
  {Landolt-B{\"o}rnstein, New Series}}},\ edited by\ \bibinfo {editor}
  {\bibfnamefont {Y.}~\bibnamefont {Shiozaki}}, \bibinfo {editor}
  {\bibfnamefont {E.}~\bibnamefont {Nakamura}}, \ and\ \bibinfo {editor}
  {\bibfnamefont {T.}~\bibnamefont {Mitsui}},\ Vol.\ \bibinfo {volume}
  {III/36a1}\ (\bibinfo  {publisher} {Springer-Verlag},\ \bibinfo {address}
  {Berlin},\ \bibinfo {year} {2001})\BibitemShut {NoStop}%
\bibitem [{\citenamefont {Duval}\ \emph {et~al.}(2007)\citenamefont {Duval},
  \citenamefont {Holtappels}, \citenamefont {Vogt}, \citenamefont
  {Pomjakushina}, \citenamefont {Conder}, \citenamefont {Stimming},\ and\
  \citenamefont {Graule}}]{duval2007}%
  \BibitemOpen
  \bibfield  {author} {\bibinfo {author} {\bibfnamefont {S.~B.~C.}\
  \bibnamefont {Duval}}, \bibinfo {author} {\bibfnamefont {P.}~\bibnamefont
  {Holtappels}}, \bibinfo {author} {\bibfnamefont {U.~F.}\ \bibnamefont
  {Vogt}}, \bibinfo {author} {\bibfnamefont {E.}~\bibnamefont {Pomjakushina}},
  \bibinfo {author} {\bibfnamefont {K.}~\bibnamefont {Conder}}, \bibinfo
  {author} {\bibfnamefont {U.}~\bibnamefont {Stimming}}, \ and\ \bibinfo
  {author} {\bibfnamefont {T.}~\bibnamefont {Graule}},\ }\href@noop {}
  {\bibfield  {journal} {\bibinfo  {journal} {Solid State Ionics}\ }\textbf
  {\bibinfo {volume} {178}},\ \bibinfo {pages} {1437} (\bibinfo {year}
  {2007})}\BibitemShut {NoStop}%
\bibitem [{\citenamefont {Babilo}, \citenamefont {Uda},\ and\ \citenamefont
  {Haile}(2007)}]{babilo2007}%
  \BibitemOpen
  \bibfield  {author} {\bibinfo {author} {\bibfnamefont {P.}~\bibnamefont
  {Babilo}}, \bibinfo {author} {\bibfnamefont {T.}~\bibnamefont {Uda}}, \ and\
  \bibinfo {author} {\bibfnamefont {S.~M.}\ \bibnamefont {Haile}},\ }\href@noop
  {} {\bibfield  {journal} {\bibinfo  {journal} {J. Mater. Res.}\ }\textbf
  {\bibinfo {volume} {22}},\ \bibinfo {pages} {1322} (\bibinfo {year}
  {2007})}\BibitemShut {NoStop}%
\bibitem [{\citenamefont {Gross}\ \emph {et~al.}(2001)\citenamefont {Gross},
  \citenamefont {Beck}, \citenamefont {Meyer}, \citenamefont {Krajewski},
  \citenamefont {Hempelmann},\ and\ \citenamefont
  {Altgeld}}]{gross2001bazr0.85me0.15o2.925}%
  \BibitemOpen
  \bibfield  {author} {\bibinfo {author} {\bibfnamefont {B.}~\bibnamefont
  {Gross}}, \bibinfo {author} {\bibfnamefont {C.}~\bibnamefont {Beck}},
  \bibinfo {author} {\bibfnamefont {F.}~\bibnamefont {Meyer}}, \bibinfo
  {author} {\bibfnamefont {T.}~\bibnamefont {Krajewski}}, \bibinfo {author}
  {\bibfnamefont {R.}~\bibnamefont {Hempelmann}}, \ and\ \bibinfo {author}
  {\bibfnamefont {H.}~\bibnamefont {Altgeld}},\ }\href {\doibase
  10.1016/S0167-2738(01)00927-4} {\bibfield  {journal} {\bibinfo  {journal}
  {Solid State Ionics}\ }\textbf {\bibinfo {volume} {145}},\ \bibinfo {pages}
  {325} (\bibinfo {year} {2001})}\BibitemShut {NoStop}%
\bibitem [{\citenamefont {Bj{\o}rheim}\ \emph {et~al.}(2010)\citenamefont
  {Bj{\o}rheim}, \citenamefont {Kuwabara}, \citenamefont {Ahmed}, \citenamefont
  {Haugsrud}, \citenamefont {St{\o}len},\ and\ \citenamefont
  {Norby}}]{bjorheim2010}%
  \BibitemOpen
  \bibfield  {author} {\bibinfo {author} {\bibfnamefont {T.~S.}\ \bibnamefont
  {Bj{\o}rheim}}, \bibinfo {author} {\bibfnamefont {A.}~\bibnamefont
  {Kuwabara}}, \bibinfo {author} {\bibfnamefont {I.}~\bibnamefont {Ahmed}},
  \bibinfo {author} {\bibfnamefont {R.}~\bibnamefont {Haugsrud}}, \bibinfo
  {author} {\bibfnamefont {S.}~\bibnamefont {St{\o}len}}, \ and\ \bibinfo
  {author} {\bibfnamefont {T.}~\bibnamefont {Norby}},\ }\href@noop {}
  {\bibfield  {journal} {\bibinfo  {journal} {Solid State Ionics}\ }\textbf
  {\bibinfo {volume} {181}},\ \bibinfo {pages} {130} (\bibinfo {year}
  {2010})}\BibitemShut {NoStop}%
\bibitem [{\citenamefont {Nyman}, \citenamefont {Bj{\"o}rketun},\ and\
  \citenamefont {Wahnstr{\"o}m}(2011)}]{nyman2011}%
  \BibitemOpen
  \bibfield  {author} {\bibinfo {author} {\bibfnamefont {B.~J.}\ \bibnamefont
  {Nyman}}, \bibinfo {author} {\bibfnamefont {M.~E.}\ \bibnamefont
  {Bj{\"o}rketun}}, \ and\ \bibinfo {author} {\bibfnamefont {G.}~\bibnamefont
  {Wahnstr{\"o}m}},\ }\href {\doibase 10.1016/j.ssi.2011.01.013} {\bibfield
  {journal} {\bibinfo  {journal} {Solid State Ionics}\ }\textbf {\bibinfo
  {volume} {189}},\ \bibinfo {pages} {19} (\bibinfo {year} {2011})}\BibitemShut
  {NoStop}%
\bibitem [{\citenamefont {Shirpour}, \citenamefont {Merkle},\ and\
  \citenamefont {Maier}()}]{shirpour????evidence}%
  \BibitemOpen
  \bibfield  {author} {\bibinfo {author} {\bibfnamefont {M.}~\bibnamefont
  {Shirpour}}, \bibinfo {author} {\bibfnamefont {R.}~\bibnamefont {Merkle}}, \
  and\ \bibinfo {author} {\bibfnamefont {J.}~\bibnamefont {Maier}},\ }\href
  {\doibase 10.1016/j.ssi.2011.09.006} {\bibfield  {journal} {\bibinfo
  {journal} {Solid State Ionics}\ }10.1016/j.ssi.2011.09.006}\BibitemShut
  {NoStop}%
\bibitem [{\citenamefont {Iguchi}, \citenamefont {Chen}, \citenamefont {Yugami},\ and\ \citenamefont
  {Kim}(2011{\natexlab{b}})}]{iguchi2011}%
  \BibitemOpen
  \bibfield  {author} {\bibinfo {author} {\bibfnamefont {F.}~\bibnamefont
  {Iguchi}}, \bibinfo {author} {\bibfnamefont {C.-T.}~\bibnamefont {Chen}}, \
  \bibinfo {author} {\bibfnamefont {H.}~\bibnamefont {Yugami}},\ and\
  \bibinfo {author} {\bibfnamefont {S.}~\bibnamefont {Kim}},}\href
  {\doibase 10.1039/c0jm00443j} {\bibfield  {journal} {\bibinfo  {journal}
  {Journal of Materials Chemistry}\ }\textbf {\bibinfo {volume} {21}},\
  \bibinfo {pages} {16517} (\bibinfo {year} {2011}{\natexlab{b}})}\BibitemShut
  {NoStop}%
\bibitem [{\citenamefont {De~Souza}, \citenamefont {Munir}, \citenamefont {Kim},\ and\ \citenamefont
  {Martin}(2011{\natexlab{b}})}]{deSouza2011}%
  \BibitemOpen
  \bibfield  {author} {\bibinfo {author} {\bibfnamefont {R.A.}~\bibnamefont
  {de Souza}}, \bibinfo {author} {\bibfnamefont {Z.A.}~\bibnamefont {Munir}}, \
  \bibinfo {author} {\bibfnamefont {S.}~\bibnamefont {Kim}},\ and\
  \bibinfo {author} {\bibfnamefont {M.}~\bibnamefont {Martin}},} {\bibfield  {journal} {\bibinfo  {journal}
  {Solid State Ionics}\ }\textbf {\bibinfo {volume} {196}},\
  \bibinfo {pages} {1} (\bibinfo {year} {2011}{\natexlab{b}})}\BibitemShut
  {NoStop}%
\end{thebibliography}
%% INSERT .bbl-file here!
%\comment{

%

%}

\comment{
%% LIST OF FIGURE CAPTIONS
\clearpage
\begin{list}{}{\leftmargin 0cm \labelwidth 1.5cm \labelsep 1cm}
\item[\bf Fig. 1] The relaxed structure of the mirror-symmetric
	$\Sigma$3 (112) [$\bar{1}$10] tilt grain boundary.  	
\\
\item[\bf Fig. 2] Segregation energies of oxygen vacancies in the
	$\Sigma$3 (112) [$\bar{1}$10] tilt grain boundary as function
	of the location of the vacancy, calculated using DFT.
\\
\item[\bf Fig. 3] Core vacancy site-concentration and potential
		barrier for vacancy segregation energies
		$E_\t{seg}^\c=\{-0.25,-0.75,-1.25\}$~eV, where the thick lines
		correspond to $-1.25$~eV. Dashed/solid lines have been used for
		the dry/hydrated case and dotted lines indicate the bound of the
		present space-charge model, i.e. $c_\V^\c>c_A/2$. The middle panel shows interior
		vacancy- and hydroxide site-concentrations for the hydrated case.
\end{list}
}
\clearpage

\end{document}